\documentclass[sigconf]{acmart}
\AtBeginDocument{%
  }
\usepackage{array}
\setcopyright{none}
\acmYear{2026}
\acmDOI{arXiv:2605.30706}
\acmConference[LIMITS '26]{Computing within Limits}{June 23--25,
  2026}{Online}
\acmISBN{}
\begin{document}
\title{Relational Aesthesis in Permacomputing Practice: Building a Solar Powered Website from Reclaimed Materials}
\author{Nadia Mariyan Smith}
\authornote{These authors contributed equally to this research.}
\email{nadia.smith@mail.utoronto.ca}
\orcid{1234-5678-9012}
\affiliation{%
  \institution{University of Toronto}
  \city{Toronto}
  \state{Ontario}
  \country{Canada}
}
\author{Nils Bonfils}
\authornotemark[1]
\email{nils.bonfils@mail.utoronto.ca}
\affiliation{%
  \institution{University of Toronto}
  \city{Toronto}
  \state{Ontario}
  \country{Canada}
}
\author{Han Qiao}
\email{h.qiao@mail.utoronto.ca}
\affiliation{%
  \institution{University of Toronto}
  \city{Toronto}
  \state{Ontario}
  \country{Canada}
}
\author{Christoph Becker}
\email{christoph.becker@utoronto.ca}
\affiliation{%
  \institution{University of Toronto}
  \city{Toronto}
  \state{Ontario}
  \country{Canada}
}
\renewcommand{\shortauthors}{Smith et al.}
\begin{abstract}
Permacomputing is a nascent concept and community of practice concerned with developing alternative computing systems grounded in principles of resilience, reuse, sufficiency, and ecological limits. However, research engaging with permacomputing remains in an early stage of development, raising concerns about whether permacomputing can move beyond reflective critique to become a meaningful alternative practice. Through a research-through-design case study, we documented our experience moving a personal website from a data centre in Texas to a self-hosted solar-powered server built from reclaimed electronics. Guided by permacomputing principles and relational aesthesis, we explore what it takes for permacomputing to reconfigure material and perceptual relations. Our findings reveal the frictions of moving away from a maximalist techno-aesthetic while attempting to re-use already existing technologies, potential ways to overcome these challenges through building a community of practice, and the transformative potential of visibilizing and visceralizing digital infrastructures to cultivate more responsible ways of relating to technology. This paper contributes to emerging research on permacomputing and its aesthetics by bringing it into dialogue with theories of non-place and relational aesthesis. Rather than functioning as a purely symbolic gesture, permacomputing practices can cultivate greater collective autonomy, agency, and responsibility in how communities engage and create meaning within digital infrastructures. In the context of socio-ecological crises and anti-colonial transformation, our research offers a situated approach to building and relating to computing technologies in the ashes of dominant technological paradigms.

\end{abstract}


\begin{teaserfigure}
  \centering
  \includegraphics[width=0.9\textwidth]{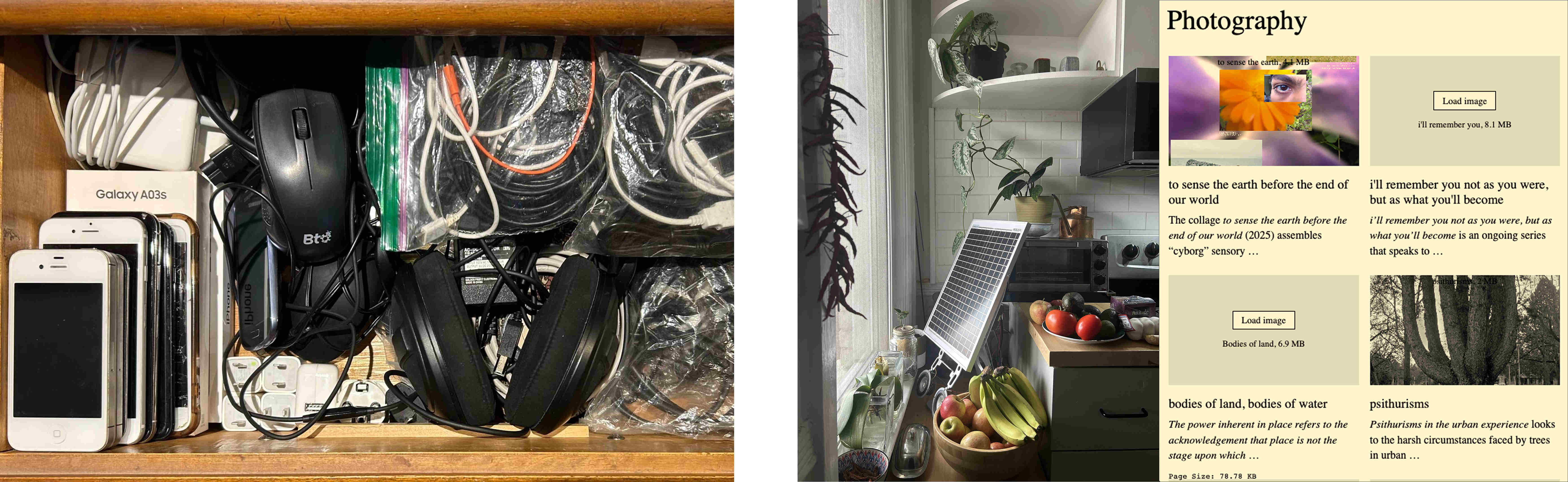}
  \caption{From the digital graveyard (left) to a solar-powered server hosting a personal website (right).}
  \Description{}
  \label{fig:teaser}
\end{teaserfigure}



\maketitle
\section{Introduction} 
\begin{quote}
“\textit{In contrast with the predominant exclusively functionalist and over-rational technocratic design models, a decolonial design would have to address more carefully and critically the realm of affect, and hence aesthesis understood in a particular way. Decolonization of the affective sphere and liberating aesthesis from the limitations of modern/colonial aesthetics is crucial for the future if we are ever to have one”} (Tlostanova, 2017).
\end{quote}
Contemporary digital infrastructures are increasingly shaped by what has been described as maximalist techno-aesthetics, aesthetics that are driven by assumptions of limitless computational capacity \cite{mansouxPermacomputingAestheticsPotential2023}. Within maximalist techno-aesthetics, digital technologies are designed to appear seamless, ubiquitous, and immaterial, obscuring the environmental conditions and labour that sustain them\cite{widdicksBreakingCornucopianParadigm2019,preistUnderstandingMitigatingEffects2016}. These conditions raise important questions about how digital technologies might be designed otherwise, particularly with how their material dependencies might be made perceptible rather than obscured. Permacomputing is one emerging response that is grounded in principles of material constraints and technological sufficiency to foster ecological responsibility.\footnote{\url{https://permacomputing.net/}} However, permacomputing remains a nascent and underexplored approach in academic research, raising questions about whether its practice can move beyond symbolic or performative gestures toward more transformative digital alternatives \cite{mansouxPermacomputingAestheticsPotential2023}.  In particular, the role of aesthetics within permacomputing deserves further exploration. 

In this paper, we examine how aesthetic practices shape how digital infrastructures produce or challenge experiences of placelessness. Decolonial design scholarship positions aesthetics as a politics of perception; a form of control that disciplines perception by prescribing what is valuable \cite{tlostanovaDecolonizingDesign2017}. The paper adopts the distinction between aesthetics as a form of representation and aesthesis as a sensory and embodied mode of perception, as understood in decolonial design scholarship \cite{tlostanovaDecolonizingDesign2017}. As we demonstrate, transforming computing practices requires not only technical change, but a reconfiguration of aesthetics; how technologies are felt, perceived, and encountered in everyday life. This aligns with decolonial critiques that technical interventions alone are insufficient if underlying logics of extraction and separability of human over nature remain intact \cite{machadodeoliveiraHospicingModernityFacing2021, escobarDesignsForThePluriverse2018}. In this context, we use "performative" to refer to aesthetic gestures that remain symbolic or representational, without reconfiguring material or perceptual relations. 

Through a research-through-design (RtD) case study, we iteratively designed and operated a solar-powered server using repurposed and reclaimed materials and reflected on the process. Our findings show how a permacomputing practice attentive to aesthesis can be articulated around three key themes: the reconfiguration of expectations of seamless technologies, the visibilization\textit{ }and visceralization of material conditions, and the role of community as a condition of technological viability. By making digital infrastructures perceptible and relational, design practices informed by permacomputing challenge dominant maximalist techno-aesthetics and open possibilities for more meaningful, empowered, and sustainable forms of computing. This paper contributes to research on the design of alternative technologies by: 1) demonstrating a web-hosting alternative using solar power and reclaimed materials; 2) extending Marc Augé's concept of non-place to digital infrastructures; 3) articulating relational aesthesis as a framework for understanding how aesthetic practices reshape technological engagement; and 4) demonstrating how a RtD approach can be used to enact and study these shifts in practice.

\section{Background}
This section introduces the theoretical and technical foundations that inform our study. We first discuss maximalist techno-aesthetics and non-place as a way of understanding placeless digital infrastructures. We then examine relational aesthesis and permacomputing, before situating our work within LIMITS scholarship on solar energy and e-waste.
\subsection{Maximalist Techno-Aesthetics and Non-Place} 
The concept of \textit{techno-aesthetics} originates in the philosophy of Gilbert Simondon, who argued that aesthetic experience emerges through interaction with technologies\cite{simondonTechnoAesthetics}.  For Simondon, aesthetics is not confined to functional appearance but unfolds through use, where both the operator and the maker experience a sensory and motoric engagement with the technical object. Building on this, Mansoux et al. describe \textit{maximalist techno-aesthetics} as the dominant aesthetic regime of contemporary computing, characterized by assumptions of ever-increasing computational power, data production, and technological complexity \cite{mansouxPermacomputingAestheticsPotential2023}.  This regime is sustained by a techno-progressive narratives the likes of the cornucopian paradigm \cite{widdicksBreakingCornucopianParadigm2019} that frame technological growth as both inevitable and desirable, promoting frictionless interfaces, seamless digital infrastructures, and continuous optimization. In doing so, maximalist techno-aesthetics encourages users to experience digital technologies as frictionless, limitless, and immaterial environments, while obscuring the material, environmental, and labour conditions that sustain them. 

To understand how this aesthetic regime shapes perception and experience, anthropologist Marc Augé's concept of \textit{non-place} provides a theoretical understanding of placeless infrastructure \cite{augeNonplacesIntroductionAnthropology1995}. Augé introduced non-places to describe the proliferation of distinct spaces in supermodernity where relational, historical, and identity-based dimensions traditionally associated with place are replaced by homogeneous environments characterized by standardized architecture, purified interiors, and the apparent erosion of social and cultural meaning. Sarah Sharma extends this concept by demonstrating that non-places are structured by biopolitical regulation that remain largely invisible\cite{sharmaBaringLifeLifestyle2009}. These spaces depend on the often invisible labour of service and maintenance workers as well as logistical infrastructures that maintain the functioning of these environments, even as they present themselves as neutral and seamless. In this paper, we draw on the concept of non-place to provide a framework for analyzing infrastructures that obscure the material and labour relations that sustain them and produce the aesthetics of placelessness.

\subsection{Relational Aesthesis and Permacomputing} 
Previous scholarship in permacomputing aesthetics has challenged the assumptions embedded within maximalist techno-aesthetics, foregrounding the role of material constraints in shaping both technological processes and cultural expressions \cite{mansouxPermacomputingAestheticsPotential2023}. Permacomputing emphasizes working within limits and allowing these constraints to shape both technical and aesthetic decisions. Understanding how these practices shape experience requires attention to perception and embodiment. Decolonial design scholarship offers a useful perspective through the concept of \textit{aesthesis,} understood as sensory perception and embodied ways of knowing \cite{tlostanovaDecolonizingDesign2017}. Tlostanova argues that modern/colonial aesthetics has historically disciplined perception by privileging abstraction and universal standards, while marginalizing relational and embodied modes of engagement. More broadly, decolonial design theorists have shown how technological design participates in the reproduction of epistemic and material hierarchies, shaping how the world is perceived and acted upon \cite{m.c.vanamsteltelCOLONIALITYMAKINGDESIGN2026}.
    
This perspective suggests that alternative computing practices must not only transform technological infrastructures but also pay close attention to the perceptual frameworks through which those infrastructures are experienced and embodied. By foregrounding sensory perception and embodied engagement with technological materials and environments, the conceptual lens of relational aesthesis may be useful in understanding how alternative computing practices can cultivate new forms of awareness regarding the material and ecological dimensions of digital infrastructures. Our paper demonstrates that relational aesthesis provides a deeper foundation for permacomputing by providing an interpretive framework through which its material practices such as reuse, repair, and working within constraints can be understood not only as technical strategies but also as perceptual interventions that reshape how computing is sensed and understood.

\subsection{Solar and E-Waste in LIMITS}
Research in the LIMITS community has framed both solar energy and electronic waste as central constraints that should actively shape computing design rather than be treated as externalities \cite{sutherlandDesignAspirationsEnergy2021,franquesaDevicesCommonsLimits2018}. Work on solar-powered and energy-harvesting systems emphasizes designing for intermittency in which services scale with available sunlight \cite{SolarProtocolExploring2022}. In particular, Low-Tech Magazine has been attracting attention due to the redesign of their online publication platform into a solar-powered website that follows their values on sustainable energy use \cite{abbingThisSolarpoweredWebsite2021}. The wider HCI literature has also engaged with solar as design constraints for websites \cite{bakhshoudehDesigningSolarInternet2025}, but also as an aid towards more relational ways of living \cite{mackeyWhatComesNoticing2025}.

In parallel, LIMITS scholarship on e-waste critiques the obsolescence of devices driven by mainstream computing practices and promotes reuse and the extension of device lifespans as key sustainability strategies, highlighting the viability of repurposing older hardware \cite{rigaudZombitronToolboxRepurposing2025}. Together, these strands converge on a design paradigm that favours sufficiency over efficiency, encouraging systems that operate within material and energetic limits by leveraging discarded devices and renewable energy sources \cite{snodgrassWindternetDesigningGridliberated2024}.

Beyond LIMITS, literature has demonstrated the importance of extending the life of smartphones and how they are often an untapped resource \cite{hazelwoodLifeExtensionElectronic2021,liSmartphoneEvolutionReuse2010,wilsonHibernatingMobilePhone2017}. Further, Switzer et al. have demonstrated that repurposing discarded smartphone "shows excellent potential for building economic and carbon-efficient systems" \cite{switzerJunkyardComputingRepurposing2023}. This provides a strong foundation for projects such as solar-powered websites hosted on reclaimed smartphones, where constraints in power availability and hardware capability are not limitations to overcome but generative conditions that shape low-impact computing infrastructures.

\section{Research-through-design}
The first three authors adopted a research-through-Design (RtD) approach, in which design practices serve as a primary mode of inquiry for generating knowledge \cite{frayling_research_1994, zimmerman_research_2007, gaver_what_2012}, to explore the research question of how designs informed by permacomputing aesthetics hold transformative potential. RtD is particularly suited to exploring emergent and entangled concepts where relevant practices are not yet stabilized and must be explored through situated making. In this project, we iteratively designed and implemented a solar-powered server and website inspired and guided by principles of permacomputing \cite{permacomputingnet_permacomputing_2025}. More specifically, our project was informed by the principles including ``care for all hardware,'' ``integrate biological and renewable resources,'' ``expose the seams,'' and ``(almost) everything has a place.'' The principle of caring for all hardware means to care for the devices that were produced from Earth's finite resources, encouraging maintenance, reuse, and extension of their lifespan. In response, we reclaimed and repurposed e-waste, including decommissioned solar panels and old mobile devices, integrating them into our system through repair and reuse. The principle of integrating biological and renewable resources guided our decision to design a solar-powered device, situating computation within the constraints of renewable energy. This reliance on solar power allowed us to reflect on assumptions of constant availability in digital infrastructures. We also engaged with the principle of exposing the seams at both hardware and web interface levels. Rather than enclosing the server within a polished container, we left its components and material visible. We also made design decisions on the website that require users to actively click on a button to load images. In doing so, processes that are typically hidden are made more visible. Finally, our project is grounded in the principle that almost everything has a place. We articulate the concept of non-place in relation to the design space of digital tools, and ask how our DIY solar-powered server can cultivate a sense of place that counters the dominant placeless and abstracted aesthetics.

Rather than optimizing toward a fixed solution, our design process involved ongoing reflection, discussions, learning, adaptation in response to material limitations, technical barriers, and shifting design goals. These moments of friction were treated as generative, revealing tensions between dominant assumptions of seamless, high-availability computing and the possibilities of an alternative future relying on low-resource infrastructures. In line with RtD traditions, the resulting artifacts and the conditions of their production function as epistemic probes \cite{bardzell_immodest_2015, giaccardi_research_2017}, enabling us to surface insights about how permacomputing and permacomputing aesthetics can be reliably enacted in practice and how alternative relational configuration between people and technology may be cultivated through design. To account for the experiential nature of this work, we incorporate a reflective practice \cite{schon_reflective_1983, godin_aspects_2014} throughout the project. The first three authors kept a journal of the building processes and our thoughts on a shared document. We attended to documenting both technical decisions and feelings and reflections articulated through ongoing communication (e.g. messages, calls, journaling). We intentionally created space for collective reflection and discussion through responding to each other's messages and holding regular meetings to exchange perspectives and surface emerging insights relevant to our research question. Although the fourth author did not participate in journaling and daily message exchanges, he engaged at key milestones of the project, bringing in his technical and research experience, providing critical feedback on the design of the artifact, evaluation procedure and research outcomes. As patterns began to stabilize, the first three authors conducted a series of longer, more focused analysis sessions in which each author systematically reviewed their notes to identify recurring themes. These themes were then discussed, compared, and iteratively refined through collective reflection amongst all authors.

\subsection{Positionality}
Our design process began as a collaboration between the first and second authors, Nadia and Nils, who are both PhD students at the University of Toronto exploring alternative technologies grounded in sustainability and social justice. 

Nadia identifies as a 'recovering user experience (UX) designer,' having been trained to develop digital interfaces aligned with maximalist techno-aesthetics. After encountering the solar-powered website of Low-Tech Magazine, they adapted its open-source template for their own site, but postponed the more technically-intensive step of building the solar-powered server. A turning point occurred when Nadia realized that their website, despite its solar aesthetic, was hosted on a data centre in Texas. This revealed 
a contradiction: the website produced what might be understood as a performative permacomputing aesthetic, adopting its visual language without its infrastructural and relational practices. This realization became the catalyst for the project, with Nils contributing technical expertise to support the development of the solar-powered server. 

Nils considers himself at ease with computers and a seasoned practitioner of self-hosting. His academic and professional background in Computer Science and Software Engineering are reflective of his lifelong interest and fascination for computing technologies. His continuous and deep engagement with computing technologies has always been coated with a particular affinity towards environmental and social sustainability. This sensitivity is what allowed him to start noticing the seams in futurist and techno-optimistic narratives. Realizing the unsustainability of current trajectories in information and computing technologies, Nils has engaged with alternative computing practices, notably those compatible with degrowth such as permacomputing.

Han, the third author, later encountered this ongoing project through conversations with Nadia and Nils, and immediately became curious about the concept and practice of permacomputing. Han has a background in Media Arts and Sciences, worked as a front-end web designer and developer for several projects, yet unaware of the underlying computer hardware and infrastructures that power web applications. During her PhD studies, she began exploring the politics of design and became increasingly aware of how dominant technological design paradigms contribute to environmental destruction. However, at times, she feels frustrated by only being able to critique and hope to materially intervene. Thus, when she learned about the ongoing project of building a solar-powered server to host websites, she became instantly interested in working together to learn about permacomputing and low-tech design approaches to digital systems.

Christoph, the fourth author, considers himself a 'recovered computer scientist' aiming to reorient computing towards just sustainabilities~\cite{beckerInsolventHowReorient2023}. From this standpoint, recognizing the recklessness of the supposedly infinite expansion of computing infrastructures instills a sense of urgency for alternative approaches to technologies 'otherwise' built on ecologically responsible foundations.

\subsection{Building Process} 
Over the course of two months, we iteratively designed and built a prototype solar-powered website with second-hand hardware and reclaimed e-waste. During this process, we consulted with mentors in our networks, and tested various hardware and software configurations (listed in Table \ref{tab:hardwareconfigurations} and \ref{tab:partslist}). A key principle guiding the development of the prototype was the reuse and repurposing of available materials. Several components were sourced from previous projects or existing devices rather than purchased new. For example, we repurposed a small solar panel previously used in a separate project that involved reclaiming discarded luxury car speakers and a laptop battery to build a solar-powered Bluetooth speaker system. Additional components included discarded smartphones, an ESP32C6 microcontroller, and an SD card module; the latter two were borrowed from teaching labs where Nadia and Nils are teaching assistants. 

    The development process was collaborative but somewhat divided across areas of expertise. Nadia primarily focused on hardware experimentation, testing different configurations, power options, and device compatibility. Nils primarily concentrated on software development and configuration, including writing and modifying code to run the web server. Han concentrated on making the website more lightweight. Instructions, successes, and failures were shared iteratively among Nadia, Nils and Han. 

    The initial design was supposed to use an ESP32C6 microcontroller as the web server but we then pivoted to repurposed mobile devices. The decision came after facing technical challenges with the ESP32 and after realizing the sustainability potential of repurposing discarded smartphones. Nadia and Nils then explored postmarketOS, a free and open-sourced Linux-based operating system developed with the goal of extending the life of consumer electronics\footnote{\url{https://postmarketos.org/}}. Although postmarketOS currently supports only a limited range of devices, it appeared to be a promising option to reclaim older smartphones and turn them into lightweight servers. After reaching out to our networks to see if anyone had any of the compatible phones listed, Nadia acquired a second-hand Samsung Galaxy S III Mini from a local seller in the Greater Toronto Area. Together with Nadia's Samsung Galaxy A03, the S III Mini proved to be incompatible with postmarketOS due to specificities in the model (North American model vs. European model).
\begin{figure}
    \centering
    \includegraphics[width=1\linewidth]{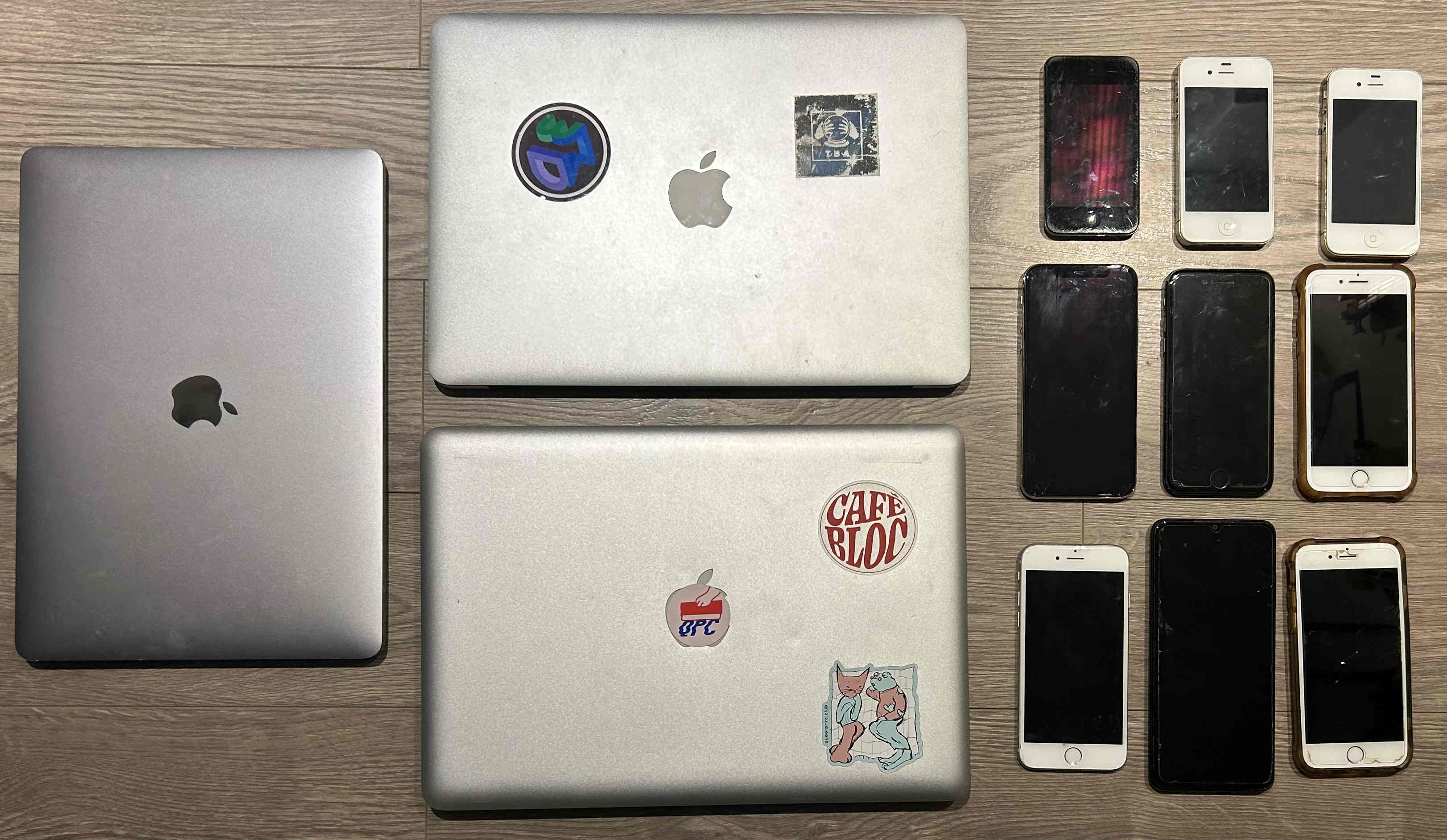}
    \caption{Maximalist computing technology found in Nadia's electronics graveyard drawer, arranged neatly on the floor.}
    \label{fig:maximalisttechnology}
\end{figure}
Our last approach was to use the Termux application, which provides an environment similar to a Linux with a terminal emulator\footnote{\url{https://termux.dev/en/}}. The only device compatible with Termux was the Samsung Galaxy A03. Through Termux, it is possible to set up remote access through an SSH server, which allows us to copy files on the device remotely and thus serves for the deployment of the website files. The website is served with the lighttpd web server\footnote{\url{https://www.lighttpd.net/}}, which is also installable on Termux. Nils first tested this configuration on his own device (OnePlus 5T) and subsequently shared the setup instructions with Nadia, who was able to replicate the setup on the Galaxy A03. Website files remained stored and edited locally on our personal machines and were copied remotely to the server through SSH for deployment. Overall, this process allowed us to evaluate the feasibility of running the server using repurposed consumer hardware while minimizing the need for newly manufactured components. 
\begin{figure}
    \centering
    \includegraphics[width=1\linewidth]{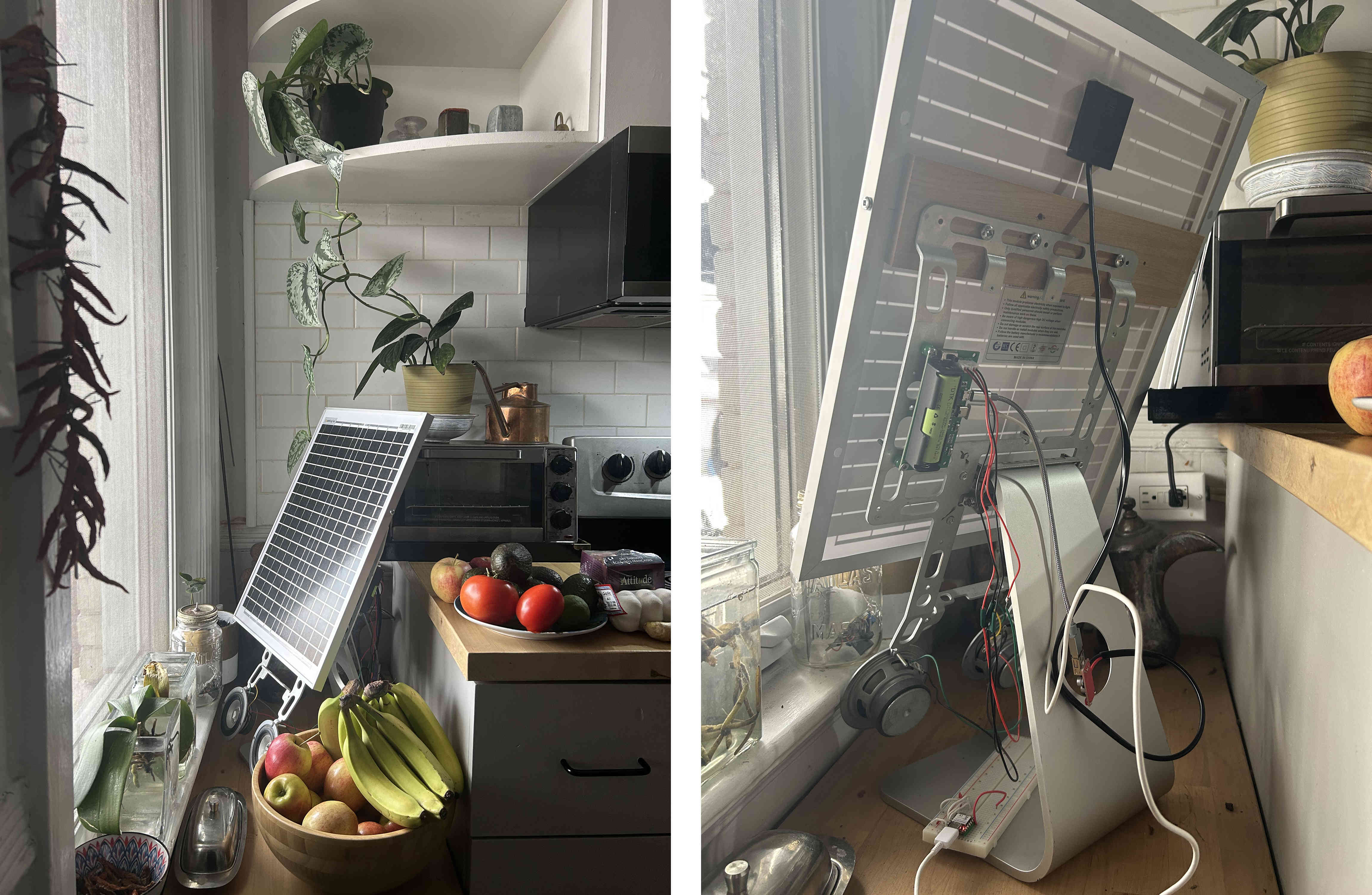}
    \caption{Iteration of the solar-powered server situated next to an east-facing window in Nadia's kitchen.}
    \label{fig:iterationofserver}
\end{figure}
\begin{figure}
    \centering
    \includegraphics[width=1\linewidth]{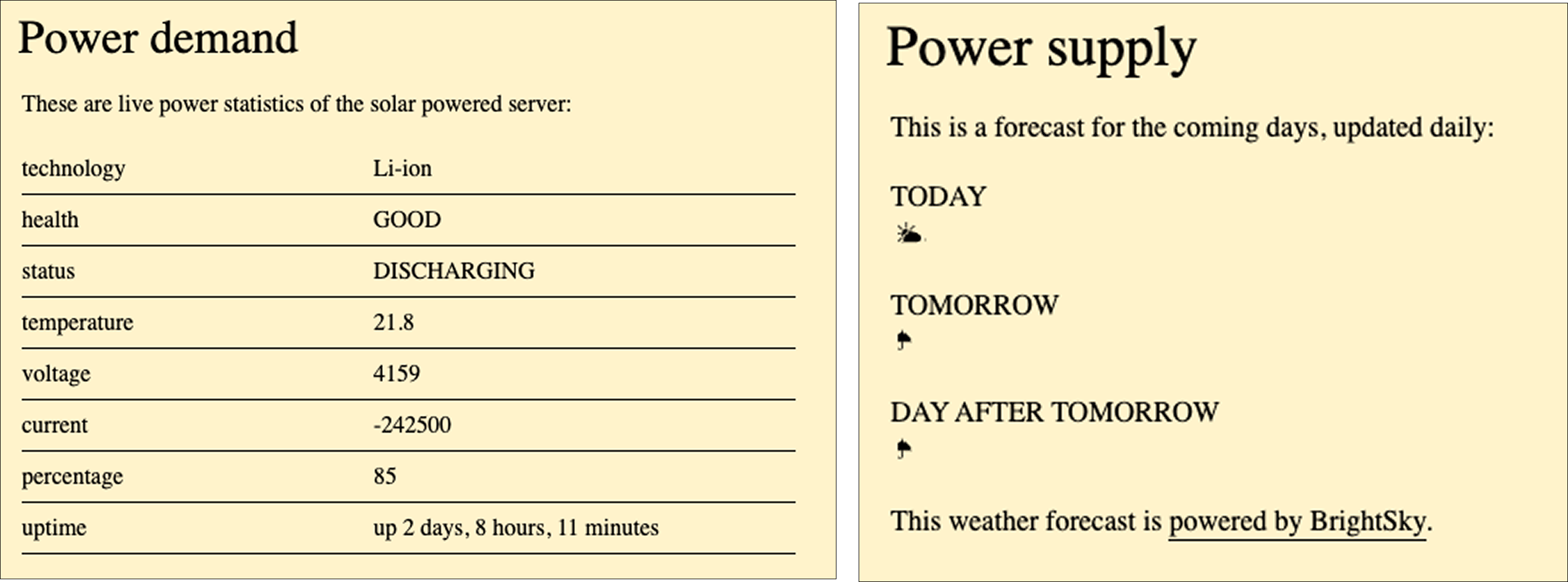}
    \caption{Screenshots of Nadia's personal website adapted from Low-Tech Magazine, including data on power demand and battery information (left) and power supply and weather information (right).}
    \label{fig:websitescreenshots}
\end{figure}  
\subsection{Setup Assessment} 
One of our goals in this project was the ability to move Nadia's personal website from GitHub pages to a more sustainable self-hosted setup powered by solar energy. To better understand the device we built, we created criteria to assess the characteristics of our prototype. These allow us to understand how successful the build was and what the priorities were in terms of technical production and output.

\paragraph{Ability to host a website on visible and tangible hardware.} In other words, whether we were able to serve a functional website from a computer that we have access to and can be maintained and cared for. This was achieved by deploying the website on our prototype and implementing the necessary network setup to make the website publicly accessible on \url{https://solar.nadiamariyan.ca}. Although repurposed devices such as the Samsung Galaxy A03 allowed for successful hosting and greater proximity to the infrastructure, they remained partially constrained by proprietary operating systems and hardware opacity. These systems made certain aspects of the infrastructure visible, such as local hosting and device-level interactions, while still obscuring others through software abstraction and platform restrictions.

\paragraph{Low operating environmental impact.} By using solar energy, we are ensuring that the environmental impact of powering our prototype remains minimal. Since the surface area of the solar panel is small (45cm x 35cm), the disturbances to the surrounding environment are fairly low and are limited to Nadia's backyard. Through our study we found that rather than being a disturbance, the solar panel actually had the potential to bolster a stronger environmental sensitivity by being a visible testimony of the dependence of our server on the sun.

\paragraph{Low embedded environmental impact.} Our privileged position as inhabitants of the abundant West means that we have access to a large amount of second-hand and discarded hardware. We acknowledge that privilege while attempting to reclaim hardware that was meant to be discarded or unused. The materials were sourced from Nadia's "electronics graveyard" and repurposed from another project. The solar panel was already second-hand in that other project and acquired from a local reseller in the Greater Toronto Area\footnote{\url{https://solarshoppingmall.com/}}.  

\paragraph{Static web hosting performance.} Nadia's personal website is static and thus does not require a lot of computational resources to serve its content. However, because the website is largely about visual arts, the content is heavy with media (e.g. high quality pictures). This requires a minimum amount of available bandwidth to serve the content at a decent speed. We did some basic traffic load testing between the website hosted on GitHub pages and the one hosted on our prototype and the results indicate that our prototype performs better in terms of static content delivery (Table \ref{tab:trafficload}). The performance is more than sufficient for our purposes and could likely allow for a dozen or more websites to be served from our prototype without degrading the performance.
\begin{table}
    \centering
    \begin{tabular}{|m{0.2\textwidth} m{0.11\textwidth} m{0.11\textwidth}|}\toprule
         \textbf{Metric}&  \textbf{GitHub Pages}& \textbf{Solar Website}\\\midrule
         Average requests per second &  241.36 & 339.07 \\
         Average data transferred per second & 1.93 MB & 2.03 MB\\
         \bottomrule
    \end{tabular}
    \caption{Traffic load testing on 16 threads, 32 concurrent connections over 30 seconds.}
    \label{tab:trafficload}
\end{table}
\paragraph{Uptime.} This was the least important of our concerns because of the nature of the personal website not requiring strong availability. We expected that the constraint of relying solely on solar energy harvesting to power our prototype would lead to downtime. However, the power draw of the smartphone combined with the power bank with insufficient power storage lead to significant downtime, with uptime ratings estimated at around 70\% from informal and non-systematic observations.
\begin{table*}
    \centering
    \begin{tabular}{|m{0.15\textwidth} m{0.1\textwidth} m{0.07\textwidth} m{0.23\textwidth} m{0.22\textwidth} m{0.1\textwidth}|}\toprule
         \textbf{Server host}&  \textbf{Software} &\textbf{Deployed}& \textbf{Considerations} &\textbf{Tangibility and Visbility}& \textbf{Power usage}\\\midrule
 Texas data center& GitHub io& Yes& Proprietary "cloud" platform &Not tangible and low visibility, infrastructure hidden in data center&unknown\\
         ESP32C6&   NodeMCU&No& Model C6 not very well supported by NodeMCU &Tangible and visibile&N/A\\
         Samsung Galaxy S III Mini& postmarketOS; remote access via Termux&No& Specific model not supported by postmarketOS; not supported by Termux &Tangible and visibile&N/A\\
         Samsung Galaxy A03&  remote access via Termux &Yes&  Not supported by postmarketOS &Tangible and visibile&low\\ \bottomrule
  
    \end{tabular}
    \caption{The hardware and software configurations that were considered viable and were tested during the build process.}
    \label{tab:hardwareconfigurations}
\end{table*}    
\begin{table}
    \centering
    \begin{tabular}{|m{0.45\textwidth}|}\toprule
         \textbf{Parts List}\\\midrule
         20W, 12V solar panel (with internal blocking diode) \\
         12V to 5V 3A USB Voltage Converter Regulator/Step Down Voltage Module \\
         5000mAH 3.7V External battery pack\\
         Samsung Galaxy A03\\ \bottomrule
    \end{tabular}
    \caption{The list of components used in our current prototype.}
    \label{tab:partslist}
\end{table}
\section{Findings}
Our findings are organized around four themes that emerged through the design and reflection process. We first describe the frictions encountered when attempting to work outside dominant technological infrastructures. We then examine the role of community in sustaining permacomputing practice, followed by how aesthetics functioned to visibilize and visceralize relations to digital infrastructure.
\subsection{Frictions reconfigure expectations of seamless technologies}  
Friction and constraint were persistent features throughout the process of building the solar-powered server, shaping both the development of the system and our experiences building it. While we each encountered friction, these experiences varied differently depending on prior technical familiarity. Nils' experience with technical systems programmed him with an understanding that frictions are part of any attempt to obtain agency and work outside of the frame that was designed. In contrast, Nadia and Han experienced these frictions more acutely. 

Throughout the building process, frictions arose from attempts to work across and outside dominant technological infrastructures. Many of the challenges encountered while experimenting with alternative software and repurposed hardware were directly related to the constraints imposed by dominant systems. This revealed how deeply these infrastructures structure what is possible. These frictions were recurring and shaped how the prototype could be developed and maintained. 

\begin{figure}
    \centering
    \includegraphics[width=1\linewidth]{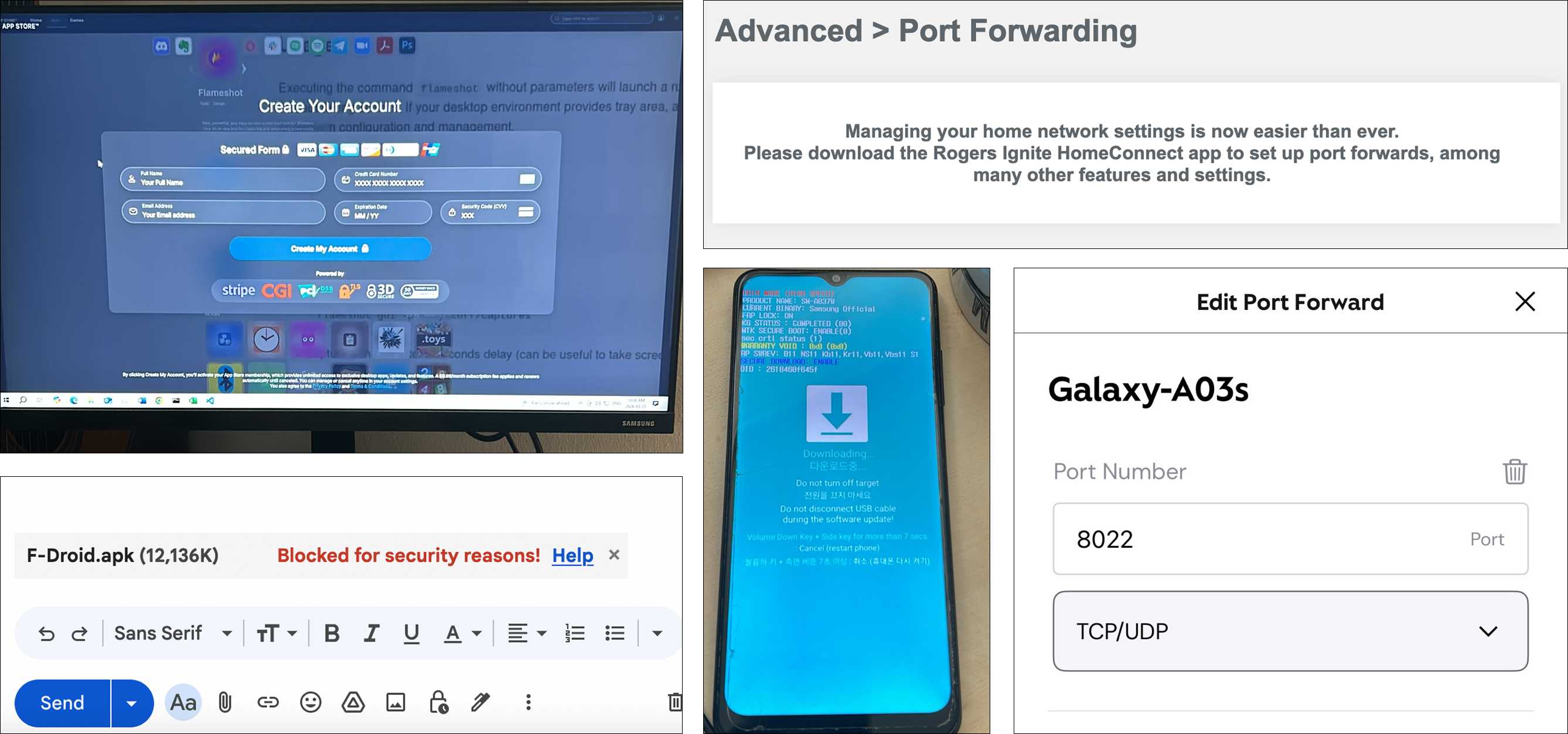}
    \caption{Frictions we encountered when engaging with dominant technology: the virus while attempting to root the smartphone (top left); router configurations requiring the Rogers Ignite HomeConnect app (top right); email with the F-Droid.apk attachment is "blocked for security reasons" (bottom left); no option to unlock the OEM on the Samsung Galaxy A03 (bottom middle); no option to specify incoming and outgoing port numbers on the Rogers Ingnite HomeConnect app (bottom right).}
    \label{fig:frictionswithdominanttech}
\end{figure}

One source of friction emerged in attempt to gain control over network infrastructure. Configuring port forwarding settings through the internet service provider required downloading their app (Figure \ref{fig:frictionswithdominanttech}). The application, however, offered only limited configuration options and did not support the necessary port forwarding capability that we needed to host the web server. Achieving the desired setup required an additional router that we sourced from our research lab's electronic drawer, where we configured a demilitarized zone (DMZ) and finally enabled the port forwarding requirements, despite increasing system complexity. 

Another source of friction was encountered through efforts to install and use alternative software on older smartphones. To begin with, attempts to install F-Droid and Termux were repeatedly met with warnings that framed these actions as unsafe. Files transferred via email were blocked for "security reasons" (Figure \ref{fig:frictionswithdominanttech}), and installation processes prompted alerts that discouraged proceeding. These interactions positioned attempts to modify device functionality as deviant, reinforcing the feeling that we are operating within tightly controlled environments. Further,  attempts to use older smartphones were further complicated by software dependencies tied to dominant ecosystems, including outdated operating systems that could no longer support applications such as F-Droid or Termux. 

More advanced attempts to modify devices, such as rooting the smartphones to gain total control over the system software, introduced additional layers of friction. Rooting required following fragmented instructions from online forums and downloading files from unverified sources. Nadia's desktop computer even got infected by a virus trying to get the necessary files for rooting the phone. Despite extended effort, rooting was unsuccessful due to manufacturer-imposed firmware locks that prevented deeper modification of the device.

These constraints extended into efforts to work with alternative operating systems. PostmarketOS, an open-source operating system designed to extend the lifespan of mobile devices, was initially explored as a potential solution. However, its limited device support posed major challenges as none of the researchers nor their personal networks owned a device that would fit our needs for the web-server and be supported by PostmarketOS. We purchased an old, used smartphone from a local seller specifically for this purpose but it was incompatible due to differences between regional hardware models.

As a result, working with alternative systems required navigating a series of cascading constraints, including sourcing compatible application versions, finding alternative file transfer methods, and attempting to modify proprietary devices. These processes were often time intensive and did not guarantee success, leading us to set aside certain approaches after extended periods of troubleshooting.

Hardware components also introduced friction, such as the external battery pack providing minimal feedback through basic indicator lights, offering little insight into system performance and battery health. The internal components of the external battery pack and the smartphone were enclosed within their casings, making it difficult to measure system behaviour directly, complicating efforts to diagnose issues such as energy retention and charge rate. 

Friction also shaped participants' emotional responses to reclaiming dominant technology. Encounters with repeated barriers produced frustration and discouragement. As Nadia noted in a Signal message to the group, "the hoops you have to jump over makes it really discouraging... so their tactic is working on me," reflecting on the perception that dominant systems actively deter attempts to gain control or understanding. Han replied "I feel the same." At the same time, frictions encountered when navigating underlying processes were seen as a positive learning experience. Nadia reflected that "maybe when there is friction, things \textit{are} going my way, because it means I get to learn more!" Rather than valuing systems that "just work," but with little control, Nadia began to value systems that could be understood and modified. But, while friction enabled learning and deeper engagement, it also introduced significant time costs and uncertainty. We often weighed the efforts required to pursue particular paths against the likelihood of success and sometimes had to stop in our tracks due to the estimated time and effort before reaching a desired outcome.
Overall, frictions primarily emerged through imposed infrastructural constraints, revealing the restricted and often opaque conditions under which dominant technologies can be accessed and modified. These findings challenge assumptions of seamlessness in dominant technology design, highlighting the time, effort, and the ongoing negotiation required to work with dominant computational systems when attempting to exercise agency beyond predefined workflows built under the guise of convenience. 

\subsection{Community of Practice}
The development process was sustained through a distributed network of collaboration, skill-sharing, and mutual aid. Knowledge was exchanged between researchers with differing backgrounds, with Nils supporting software configuration, Nadia focusing on hardware experimentation, and Han implementing energy saving designs for Nadia's personal website to be hosted on the server. Examples, instructions, code, and troubleshooting strategies were shared constantly and iteratively throughout the building process. Nadia and Han relied on Nils' technical experience to navigate unfamiliar hardware and software concepts, and Nils, in return, learned about "slowing down" and breaking down concepts, and exploring ways to communicate and make complex concepts accessible to people coming from various backgrounds. This collaborative dynamic extended beyond the core research team. Mentors, peers, and online forums contributed ideas, knowledge, and material resources, including electronic components and guidance on hardware assembly. These interactions played a critical role in enabling progress. 

Having a community that supports each other also allowed members of the community to develop confidence needed to engage with unfamiliar domains, despite the frictions and risks described in the previous section. Nadia felt more assured experimenting with booting their phone when Nils was available to provide guidance. Han described gaining confidence through seeing how Nadia, who shared similar non-technical background, actively working with the systems. 

The community was further sustained through sharing a practice of making. Working together on the server and website created a shared experience. Without this ongoing practice, discussions of permacomputing would have remained abstract or speculative. Throughout our reflection sessions, Nadia and Han repeatedly emphasized the hands-on engagement, such as connecting to the server remotely via ssh and downloading a video recording of their meeting hosted on this server, constituted their most meaningful learning experience. Nils also noted that observing how Nadia and Han navigated these processes helped him understand possibilities for how permacomputing could be taken up in practice. Such hands-on involvement through experimentation, breakdowns, and iterative problem-solving made it possible to understand the frictions as well as the conditions for such a nascent concept to become feasible and reliable.

Finally, encounters with others outside of the project further illustrate how the prototype functioned as a site of shared learning. When Nadia's friends encountered the device in Nadia's kitchen, they often asked questions such as "what is a server?" and "how does this device work?", which led to discussions about digital infrastructure and sparked interest in wanting to build their own. Notably, Han who also had similar questions when Nadia shared the project idea with her, later became directly involved in this project. 

\subsection{Aesthetics for visibilizing hidden infrastructure}
A key role of aesthetics in this project was to make visible the material conditions of digital systems that are typically obscured in maximalist techno-aesthetics. 

The physical device was assembled from repurposed and reclaimed materials, including a solar panel, wood, a Mac monitor stand, speakers from a discarded luxury car, a step-down voltage converter, a rechargeable battery pack, cables, and a smartphone. It's internal structure, wires, components, and connections were intentionally exposed rather than concealed, allowing the device to be read and explained through direct observation. Built-in affordances on the step-down voltage converter and the rechargeable battery pack revealed system states by displaying LED lights corresponding to current flow and charge levels, respectively. When describing the project to friends, the device itself or photos of the device would be used to help describe the server and the interconnected components.

The visibility of infrastructure extended to the interface design of the website hosted on the server. The site incorporates real-time system information, features that made the system's conditions and constraints visible. These include server status, location, battery levels, weather conditions, inspired by Low-Tech Magazine's website \footnote{https://solar.lowtechmagazine.com/} (Figure \ref{fig:websitescreenshots}). Interactions such as tap-to-load images alongside image sizes (Figure \ref{fig:teaser}) introduce a constraint that draws attention to resource use, inspired by Permacomputing as a Practice for Digital Graphic Design \footnote{https://permacomputingasapracticefordigitalgraphicdesign.com/}. 

As noted in our reflection on the RtD process, these material and interface elements "closed the gap," or reduced the distance, between hardware and software, making the system legible as a set of interdependent relations. By making the dependencies of the prototype visible, aesthetics functioned as a means of revealing the infrastructural conditions of the technology, transforming the system from a non-place technology to one that felt more place\textit{full.}

\subsection{Aesthetics for visceralizing relations to technology}
Beyond visibility, aesthetics also played a role in shaping how we felt, related to, and engaged with the system, producing what we described as a visceral relationship to the technology. While encounters with technical challenges often produced frustration, these experiences were gradually reframed as opportunities for deeper understanding and learning. Moments of progress such as successfully configuring remote access, were experienced as meaningful and rewarding, contributing to a growing sense of confidence and capability. Joy and intrigue were often expressed in these moments. For example, both Nadia and Han described the experience of accessing battery information remotely as "very cool" and "exciting." The direct access to system information that is typically hidden enabled a more immediate understanding of the system's operation, connecting abstract concepts to tangible resource conditions. 

Closing the gap between hardware and software was not only visually understood but viscerally felt. For example, Nadia described a moment of realization when Nils shared a file remotely through the server: "omg the video is sitting on the device in my kitchen right now and we can all just grab it," highlighting a shift from passive consumption to active participation. Another example is when Nadia described not needing to know the model number of the Samsung device until wanting to understand the device to have more autonomy over it. The Samsung smartphone became something to be understood, modified, and maintained rather than passively used.

\begin{figure}
    \centering
    \includegraphics[width=1\linewidth]{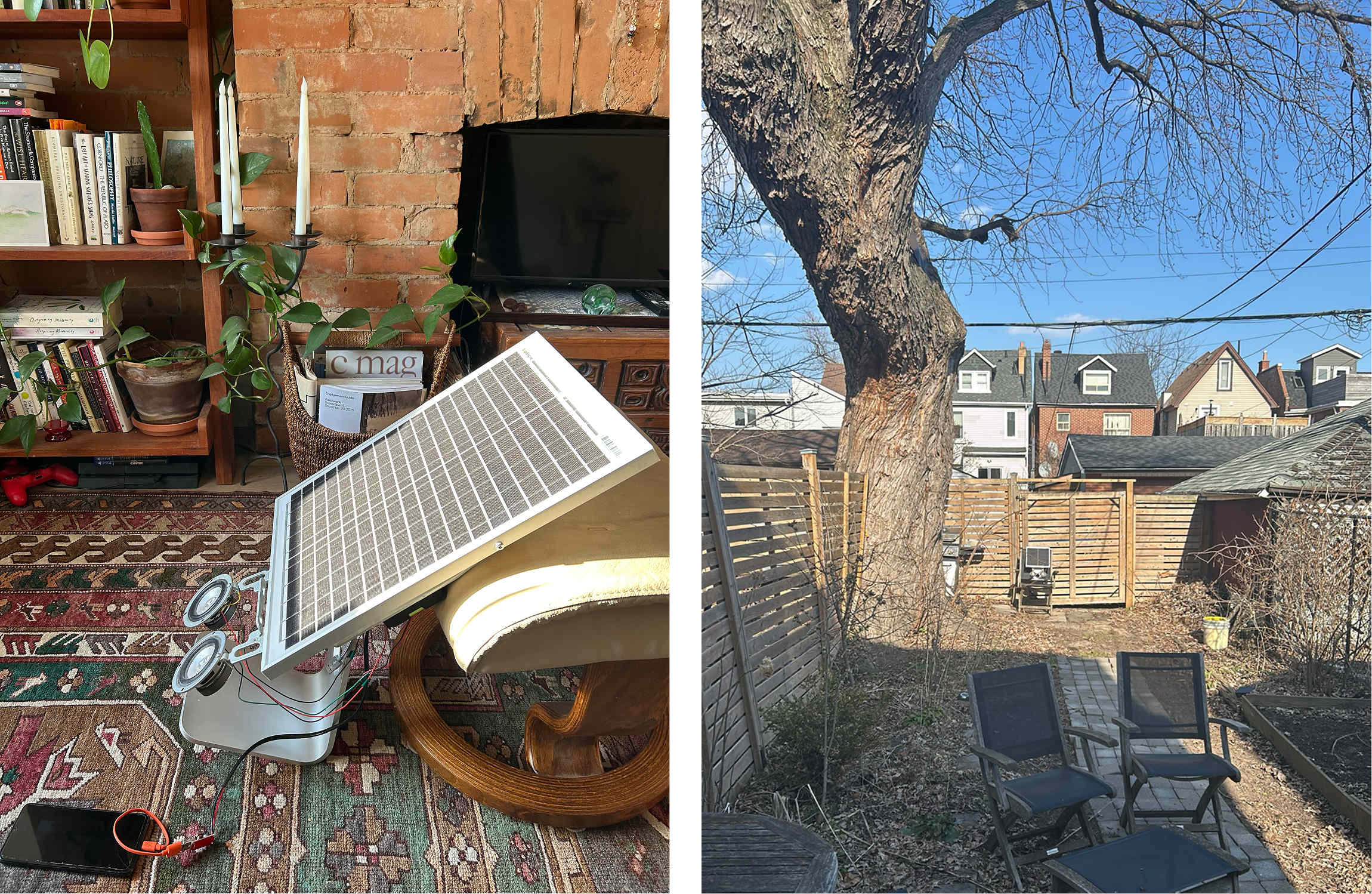}
    \caption{ The server is positioned and angled in a way to get direct sunlight from Nadia's living room (left) and at the very back of Nadia's backyard (right).}
    \label{fig:serverplacements}
\end{figure}

    Assembling the server also introduced new forms of engagement tied to its material conditions. The building process involved pivoting between hardware to host the server, battery sizes and additions, and trials of different configurations to find a setup that would sufficiently charge a host for our server, all while using reclaimed materials that we had access to. In an earlier iteration of the prototype, maintaining a healthy battery charge involved Nadia physically repositioning the device throughout the day in response to sunlight availability; placing it by an east-facing window in the morning, moving it outdoors in their backyard during peak sunlight in the afternoon, and relocating it to the west-facing window in the late afternoon (Figure \ref{fig:serverplacements}). This process depends on the weather conditions and Nadia's presence at home. The components we had access to brought forth frictions that made Nadia more in tune with the device; when the external or smartphone battery drained, the devices needed to be turned back on during hours of sunlight in order for the system to start harvesting energy and for the server to turn on. 
    The device also carried stories: components reused from previous projects and collaborations contributed to a narrative that extended beyond its immediate function as a server. When describing the device to friends, Nadia found that it was difficult to separate the object from its history, describing attachment to both its aesthetic form with the bluetooth speakers and the labour involved in its creation. They would often describe the build that involved using reclaimed luxury car speakers to create a solar-powered Bluetooth speaker that they made in a workshop with mentor Brian Sutherland, as well as the process of the subsequent transformation of the build into the solar-powered server with Nils and Han. Nadia expressed to Nils that they "love that it is a unique object" and that it "enables discussion and reflection" when friends and family visit their home. 

    Through these experiences, aesthetics not only operated at the level of appearance, but also as an embodied process. In the process of aligning the aesthetics of the server to show its placefullness, the prototype transformed how we related to technology, to varying degrees, from distant and abstract systems to situated, meaningful, and felt engagements.

\section{Discussion}
In this section, we connect our findings to broader conversations in HCI, permacomputing, and decolonial design. We first discuss how the prototype challenged experiences of placelessness and meaninglessness in digital infrastructures, then examine community as a condition for permacomputing practice, and finally reflect on how aesthetics can move permacomputing beyond performative gestures toward agency, responsibility, and care.
\subsection{Cultivating Meaning in a Placeless Digital Graveyard}
Our findings show that the experience of placelessness in digital systems is actively produced through the aesthetics of dominant technologies. As outlined in section 2.1, maximalist techno-aesthetics privileges seamlessness, abstraction, and ubiquity, removing any signifiers of the material and labour conditions on which dominant digital systems depend. In practice, this manifested as infrastructures that concealed energy use, such as the data centre in Texas hosting Nadia's GitHub Pages site, and restricted access to underlying processes through proprietary hardware and software, including the Rogers network router and the Samsung smartphones with experimented with. These conditions produce an experience of digital systems as detached from their contexts of operation, aligning with Sharma's account of non-place as environments where the material, labour, and political relations that sustain them are obscured, while still actively organizing how people engage with and move through these systems.

In contrast, the solar-powered server disrupted this condition of placelessness by making its dependencies perceptible and unavoidable. Its operation required ongoing attention to sunlight availability, device placement, system behaviour, and available material resources, introducing forms of visibilization that reconnected digital technology to its environmental and material constraints. The system demanded situated engagement, making its conditions of operation both visible and consequential. 

The solar-powered server not only rendered infrastructure visible, but also demanded physical and perceptual engagement, enabling a visceral experience of the system as contingent and requiring ongoing attention. Drawing on relational aesthesis as outlined in section 2.2, this reconfiguration can be understood as a transformation in how digital systems are sensed and interpreted. The website was no longer experienced as placeless, but as embedded within specific contexts that shaped its operation and meaning.

Importantly, this reconfiguration altered how meaning is produced. In dominant technological systems, as in Augé non-place, meaning is detached from the conditions of technological operation. In contrast, our prototype cultivated meaning through the building process, material conditions, and ongoing interaction. The device accumulated significance and personal attachment through its history of assembly and the labour involved in its upkeep. As described in the findings, even discarded devices reclaimed from electronics graveyards were reactivated and took on new significance, highlighting how meaning in digital systems can be cultivated through the conditions under which it is engaged rather than inherent to the object itself. 

Meaning was reinforced through social interactions. The physical presence of the device in Nadia's home enabled in-situ encounters with friends and visitors, where the object became a site of storytelling, knowledge-mobilization, and shared reflection. Its functional aesthetics comprised of its visible components and unique configuration invited discussion and curiosity, embedding the system within a broader network of relationships. In this sense, meaning emerged through collective interpretation and interaction. 

Taken together, these findings suggest that cultivating meaning in digital systems involves designing the aesthetics of the digital systems to more accurately situate them within their material and labour relations. By foregrounding these dependencies, our solar-powered server made from reclaimed materials challenges the perceptual regime of maximalist techno-aesthetics and demonstrates how technology designers can and should re-embed digital infrastructure within lived and situated contexts. 

\subsection{It Takes a Community: Collective Conditions for Permacomputing}
Our findings show that the viability of permacomputing practices depends on a networked community. Working with reclaimed maximalist technologies revealed that trial and error are central to achieving a minimally functional system. The project points towards the importance of community as infrastructure, with access to materials and knowledge emerging as critical factors in sustaining the system. This finding aligns with permacomputing as a community of practice, but our insights demonstrate how this collectivity operates as a practical condition for system viability. We drew on personal networks to source components and online forums to seek advice, highlighting how shared resources and expertise contribute to robustness. The idea that "it takes a community" became increasingly salient, as the availability of unused electronics in personal "electronics graveyards" and collective expertise significantly improved the robustness of the system. 

At the same time, participation in the project was not uniform, but emerged through situated forms of engagement. Each of us developed distinct relationships to the project, shaped by our positionality and modes of engagement. For Han, the project was meaningful as a site of community formation and learning, where knowledge was shared and developed collectively. Nadia's relationship with the project was shaped by a strong connection to its material and aesthetic dimensions, particularly through the processes of visibilizing and visceralizing the infrastructure. Having physically built the device and maintained it within their home, Nadia experienced the prototype as a meaningful object embedded in personal history and everyday life. Nils approached this project with a particular interest in how individuals with relatively limited technical backgrounds navigate such systems. His engagement centered on the process of learning and how best to support efforts to reclaim and reuse maximalist technology. These differing perspectives illustrate the multiple entry points for individuals with varying levels of technical experience to engage, learn, and contribute over time. 

This insight aligns with existing communities of practices such as Solar Protocol \footnote{https://solarprotocol.net/}, a distributed network of solar-powered servers that collectively host web content. Solar Protocol achieves reliability through distributed hosting where multiple nodes compensate for variability in local conditions for solar energy availability. This model demonstrates how robustness in alternative computing can emerge through collective organization. In such contexts, robustness is a social achievement, emerging from shared resources, mutual-aid, and ongoing collaboration. Reframing robustness in this way shifts the focus from individual efforts to cultivating collective conditions in which such systems can be created and sustained. 

\subsection{Moving Beyond Performativity through Agency, Responsibility, and Care}
While prior work has raised concerns that permacomputing risks being performative \cite{mansouxPermacomputingAestheticsPotential2023}, our findings suggest that its transformative potential lies in how its aesthetic practices are materially and perceptually enacted. Here, we understand "performative" as aesthetic gestures that remain representational or symbolic, without reconfiguring the underlying material and relational conditions of dominant technological systems. In contrast, we show that when aesthetics operate through embodied engagement and material practice, what Tlostanova conceptualizes as relational aesthesis \cite{tlostanovaDecolonizingDesign2017}, they can produce meaningful transformations in how technologies are experienced and related to. 

Our findings suggest that engaging with the solar-powered server did not only provide us with greater control over our digital system, but also reconfigured our sense of agency toward the system. In dominant digital infrastructures that we are entwined in, we noticed that we engaged as passive consumers, interacting through interfaces that conceal underlying processes and dependencies. As discussed in Section 4.1, agency within these systems is largely constrained to predefined interactions, while the material conditions remain outside of user awareness and responsibility. In contrast, the solar-powered server afforded a different mode of engagement by requiring direct interaction with and intervention in the system's operation. We monitored environmental conditions for sunlight availability, adjusted device placement accordingly, and responded to fluctuations in system performance with attempting different configurations. Through this process, the system was no longer experienced as something that "just works," but as contingent on ongoing attention and care. 

Crucially, this shift in agency was inseparable from an increased sense of responsibility, which emerged from awareness of system dependencies and from direct involvement in its making and upkeep. The time and labour invested in constructing and maintaining the system fostered a form of attachment and shaped decision-making. For example, decisions about modifying or replacing components were not made solely on the basis of efficiency or optimization. In a pivotal moment in our project, we debated on whether we should move forward with using the ESP32C6 MCU as it afforded more visibility over its functionality than a reclaimed smartphone. In another instance, Nadia chose not to remove the reclaimed luxury car speakers that were already integrated into the structure of the prototype, despite their lack of direct functional relevance to the server. The speakers were retained because they embodied the history of the device's construction originating from a prior project, and carried the visible traces of experimentation and collaboration. The device was thus encountered as a situated artifact with a history and an identity. Responsibility also emerged through the system's dependence on environmental conditions. As we became more aware of the need for more sunlight on the solar panel, we also became aware of the consequences of our inaction. Moments such as missed sunlight due to Nadia not getting out of bed early enough to turn on the empty battery to enable charging, or Nadia missing an opportunity to put the panel in a sunnier spot, highlighted how system performance was directly tied to our involvement. These experiences made clear that the system's operation was contingent on sustained engagement, reinforcing a sense of responsibility grounded in everyday practice. In this sense, agency manifested as a relational condition that required effort and commitment. 

\section{Conclusion}
Permacomputing has been proposed as an alternative computing practice that foregrounds material constraints and technological sufficiency to foster ecological responsibility. However, as a nascent community of practice, it has room to elaborate on its transformative potential. Through RtD, this paper examined what it takes to enact permacomputing in practice by migrating a personal website hosted on GitHub Pages to a solar-powered server built from reclaimed materials. Our findings suggest that the distinction between performative and transformative practices in permacomputing hinges on how aesthetics are enacted. When aesthetics remain at the level of representation, they risk reinforcing symbolic engagement without altering underlying relations. However, when aesthetics are embedded in material practice and perceptual experience, they can reconfigure how we relate to technologies, fostering greater agency, responsibility, and care for our digital technologies.
At the same time, we acknowledge that permacomputing remains entangled within dominant technological systems and faces ongoing challenges related to reliability, accessibility, and scalability. Future work should explore how distributed approaches the likes of the Solar Protocol or contributions to open-source operating system initiatives like PostmarketOS, can support more robust and accessible implementations of alternative computing practices. 
In sum, our paper demonstrates the transformative potential of permacomputing lies in its ability to mobilize aesthetics not as surface expression, but as a means of reshaping engagement with the conditions of computation itself. 


  

\begin{acks}
We would like to thank Brian Sutherland for his mentorship and his workshops on low-carbon electronics-making which were foundational to the development of the prototype. Nadia would also like to extend thanks to Paul Kemp, whose decades of experience teaching electrical engineering provided support in developing foundational knowledge for this project. We also thank the Just Sustainability Design lab for lending us the router used in our setup. We are grateful to Nadia's partner Ben for sharing their kitchen and living space as a site of experimentation and for contributing to the pool of unused electronic materials. Finally, we thank our broader networks for responding to ongoing calls for unused electronics. 
\end{acks}

\bibliographystyle{ACM-Reference-Format}
\bibliography{base}



\end{document}